\newif\ifarxiv
\arxivtrue 

\ifarxiv
\documentclass{article}
\else
\documentclass[aps,pra,twocolumn,showpacs,longbibliography,groupedaddress]{revtex4-2}
\fi

\usepackage{amsmath,amsfonts,amssymb}
\usepackage{graphicx}
\newcommand{\ket}[1]{|#1\langle}

\begin{document}

\title{Photon number resolving detectors as evidence for the corpuscular nature of light}
\ifarxiv
\author{Morgan C. Williamson, Gabriel D. Ko, and Brian R. La Cour}
\else
\author{Morgan C. Williamson}
  \email{morgan.williamson@arlut.utexas.edu}
  \affiliation{Applied Research Laboratories, 
               The University of Texas at Austin, 
               P.O. Box 8029, 
               Austin, TX 78713-8029}
\author{Gabriel D. Ko}
  \affiliation{Applied Research Laboratories, 
               The University of Texas at Austin, 
               P.O. Box 8029, 
               Austin, TX 78713-8029}
\author{Brian R. La Cour}
  \email{blacour@arlut.utexas.edu}
  \affiliation{Applied Research Laboratories, 
               The University of Texas at Austin, 
               P.O. Box 8029, 
               Austin, TX 78713-8029}
\fi

\date{\today}

\ifarxiv
\maketitle
\fi

\begin{abstract}
We consider and test the influence of a variable coincidence window on photon number statistics using a beam splitter network based quasi-photon number resolving detector. We reveal the commonly insufficient signal-to-noise ratio (SNR) of a survey of PNR detectors. The origin of the SNR deficiency mainly concerns coincidence windows that are not sufficiently brief compared to the light source coherence time. In order to illustrate this blindspot we configured a beamsplitter tree-based quasi-PNR detector with a similarly insufficient SNR and revealed a marker of concern: changing the photon statistics at will by varying the coincidence window via post-processing. We also reproduce this anomalous and concerning behavior using a model of a quasi-PNR detector that avoids any reliance on the presumption of light particles, instead remaining consistent with a wave picture of light.  We conclude that most PNR detectors are not, in and of themselves, indicative of the corpuscular nature of light.
\end{abstract}

\ifarxiv
\else
\maketitle
\fi


\section{Introduction}

Photon number resolving (PNR) detectors are an important enabling capability for photonics-based quantum computing, sensing, and communication.  While traditional threshold detectors indicate with a ``click'' only the presence or absence of one or more photons, a PNR detector indicates the \emph{number} of photons detected, up to a device-dependent upper limit.  PNR detectors take a variety of forms, from multiplexed single photon detectors to solid-state and superconducting devices producing multimodal signals \cite{Chunnilall2014,Provaznik2020}.  

On the face of it, PNR detectors would seem to provide direct evidence for the discrete and corpuscular nature of light.  Yet it is also well known that light exhibits both particle-like and wave-like behavior, and the reconciliations of these two views has long been an outstanding issue for the foundations of quantum mechanics.  The photoelectric effect, the basic mechanism underlying all photodetection, provided the first evidence for the corpuscular nature of light, yet Lamb and Scully noted years ago that a wave picture can be maintained in a semiclassical treatment of the effect \cite{LambandScully1968}.  Thus, the photoelectric effect alone does not \emph{compel} us to adopt a corpuscular view of light, even if it does simplify our description.  Similar observations have been made regarding the Compton effect \cite{Frantz1965}.

The nonlocal nature of some PNR detections would also seem to suggest a corpuscular interpretation.  Einstein first highlighted this effect in his debate with Bohr during the 1927 Solvay conference, where he considered the wavefunction of a single photon impinging on a distant screen yet registering a detection at only one point \cite{Solvay1927}.  A similar phenomenon occurs in PNR beam splitter networks when, say, weak coherent light produces a detection in just one mode.  From a purely wave picture of light, this may seem like rather odd behavior, yet it is not difficult to imagine how such phenomena could arise.  A classical description of weak coherent light, for example, in a beam splitter network includes an independent vacuum noise term at each output mode.  If one supposes that a local amplitude above a given threshold corresponds to a detection in that mode, then the conditional probability of a single detection, given that any detections at all occur, can actually be quite high.  The apparent nonlocality, in this case, is a mere artifact of statistics.  Similarly, coincident multi-mode detections in this scheme would mimic a multi-photon detection event.  It is not difficult to imagine other nonlinear transfer functions that would mimic the behavior of multi-modal PNR devices.  Thus, these features alone are not inconsistent with a wave picture of light.

As can be seen, the presence of noise can mimic the behavior of multiphoton events, and this effect will only be amplified as the number of detection modes is increased.  Two features could potentially combat this effect: a higher signal-to-noise ratio (SNR) and a smaller coincidence window.  A high SNR is needed to overcome the ambiguity of false detections, but too high of an SNR (achieved by increasing count rate) will saturate the limits of the PNR device.  Therefore, a sufficient, intermediate value is needed.  In addition, many PNR devices lack the specification of a coincidence window over which multiple detections are associated.  This may be in the form of an explicit time window, in terms of resolvable time stamps, or an implicit window defined by the spatio-temporal integration of the device.  A coincidence window at or below the coherence time of the light source should minimize false detections and provide a consistent photon number distribution.  If this window is increased and the SNR is too low, then an increase in the observed mean photon number would be expected due to an increase in false detections.  The intrinsic dependency of the photon number distribution to the detailed properties of the PNR detector in this regime calls their validity into question.

To investigate these effects, we performed an experiment using weak coherent light and a beamsplitter-based PNR detector with a low SNR in which we varied the coincidence window in post processing.  We hypothesize that with low SNR a change in the coincidence window will, in fact, cause a change to the resultant photon distribution.  The arbitrary control of output photon statistics simply by varying the coincidence window would, we believe, indicate anomalous behavior attributable to the low SNR of many PNR detectors.  To explore whether such behavior could be reproduced by a classical model, we investigated a wave-based model of light designed to replicate the general behavior of the experimental results.  The extent to which this model agrees with our experiment indicates that a corpuscular inference is not needed.  By extrapolation, PNR detectors with similarly insufficient SNR should also be regarded as not indicative of the corpuscular nature of light.

The outline of our paper is as follows:  In Secs. \ref{Experimental Setup} and \ref{Analysis Procedure} we describe our experimental setup and analysis procedure, respectively.  In Sec. \ref{Results} we compare our results against those of a simple classical model.  We discuss the broader implications of this work and compare to other PNR detectors in Sec. \ref{Discussion} , and in Sec. \ref{Conclusion} we summarize our conclusions.


\section{Experimental Setup} \label{Experimental Setup}

The PNR detector used in this work consists of a multiplexed beamsplitter tree-based network with three 50:50 beamsplitters (Thorlabs 	
CCM5-BS017) supplying four independent single-photon detectors (S-fifteen Si-APD).  A fiber-based laser diode (Thorlabs MCLS1 with ss-d6-6-785-50 diode) supplies the input coherent light at 778 nm with a coherence time of 2 ps.  An in-line fiber variable attenuator (Thorlabs VOA780-APC), fiber coupling (Thorlabs PAF2-2B), and a set of neutral density filters (Thorlabs NEK01) provide coupling to free space and attenuation.  Timing electronics (S-fifteen TDC1) provided timestamps of all detection events among the four detectors, with a timestamp step of 1 ns and a timing resolution of 2 ns.  The software-defined coincidence window was swept from 20 $\mu$s down to 10 ns in 500-ns steps.  Power was measured using a free-space power meter (Newport 843-R) that could be positioned to intersect the optical axis before the attenuation stack.

\begin{figure}[ht]
\centering
\includegraphics[width=\columnwidth]{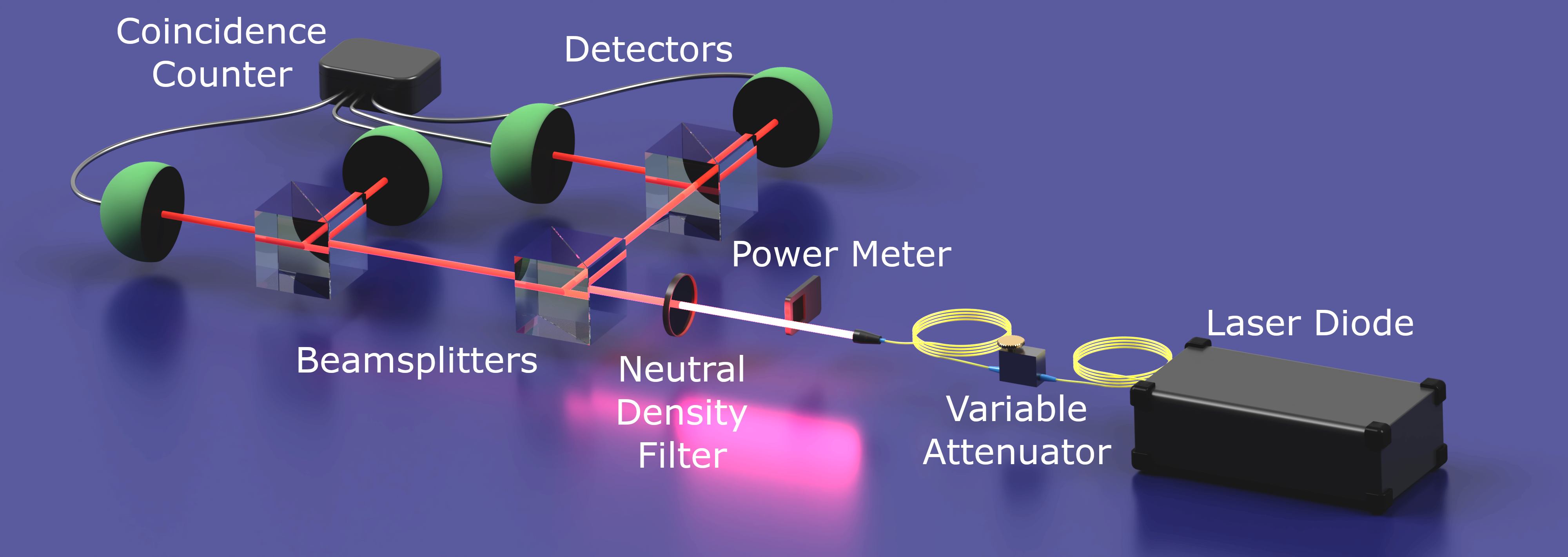}
\caption{(Color online) The experimental setup consists of four single-photon detectors, three 50/50 beamsplitters, a neutral density filter stack, a laser diode, a power meter, a variable attenuator, and a coincidence counter.  The power meter was inserted at times to intercept the beam just prior to the neutral density filter stack.}
\label{fig:setup}
\end{figure}

The single-photon detectors used for this work are of an avalanche photodiode design and are passively quenched with a dead time of about 2 $\mu$s.  They have stated nominal efficiencies of about 50\%.  The efficiencies were estimated by the manufacturer by comparing each detector to a reference detector, itself having been tested using a traceable power meter.  The detectors are also estimated as having a dark count rate of about 300 counts per second (cps).

One important complication that surfaces when incorporating multiple SPDs in a system is the relative balance of counts between detectors.  Although an exact balance is not needed, one must ensure that the amount of power received by each detector guarantees that the dynamic range of each detector largely coincides with that of the others.  This is necessary to prevent a situation in which one detector is operating in the well-behaved linear regime while another detector is either in saturation or near the dark-count regime.  In our setup this was arranged by maximizing the detector counts of all the detectors at a modest power level of 0.11 nW, isolating the detector with the lowest maximum counts, then compensating the other three detectors by slightly misaligning the fiber coupling.  This resulted in a relative weighting of counts between the detectors, which is summarized in Table \ref{tbl:detector_counts}. 

\begin{table}[ht]
\begin{tabular}{cc}
\hline
Detector \; & \; Count Rate (cps) \\
\hline
D1 & $63200 \pm 430$ \\
D2 & $55000 \pm 440$ \\
D3 & $59800 \pm 480$ \\
D4 & $61800 \pm 460$ \\
\hline
\end{tabular}
\caption{Single-photon count rates measured at 0.11 nW, used for detector balancing. The labels D1, \ldots, D4 indicate detectors 1 through 4, respectively.}
\label{tbl:detector_counts}
\end{table}
	
The intentional misalignment of fiber coupling is generally undesirable due to the lowered system detection efficiency.  It is important however to note that we are not attempting to conduct the particularly difficult and metrologically traceable system detection efficiency measurement in which minimizing losses is paramount.  Instead, we employ the power meter simply as a proportionality monitor for the intensity level. That being said, the model used herein, and detailed below, to convert click statistics to photon statistics is capable of accounting for not only unique detector efficiencies, but also unbalanced branches of the beamsplitter tree, including variances in count rates of 15\% as in this case. The neutral density filter stack used, which consisted of filters with optical densities of 2, 1, 0.6, and 0.4, produced a total attenuation factor of $962 \pm 50$. The attenuation factor is offset from the nominal value of $10^4$ due to the specific wavelength-dependence of the filters.

\begin{figure}[ht]
\centering
\includegraphics[width=\columnwidth]{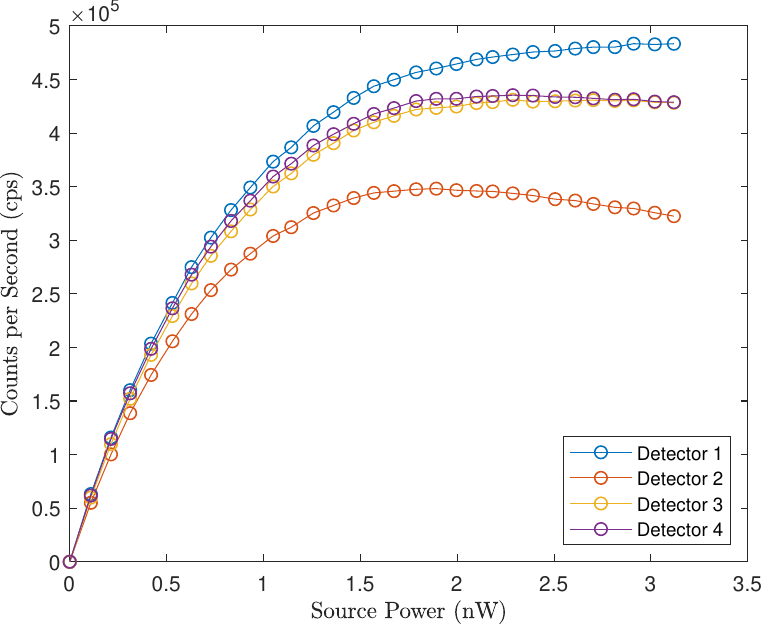}
\caption{(Color online) In-situ data displaying detector count rates as a function of source power.  Signs of detector saturation are visible throughout the dynamic range of the detectors but dominate at source powers over 2 nW.}
\label{fig:saturation}
\end{figure}

A measured laser power level of 1.9 $\mu$W before attenuation through the NDF stack (and 2 nW after attenuation) induced the onset of detector saturation, as shown in Fig.\ \ref{fig:saturation}.  As such, we chose to take measurements at 21 different power levels, in equal increments from 2 nW down to levels near the noise floor or dark-count regime.  The dark count regime measurements were taken with no change in the optical setup except the closing of the in-line fiber attenuator such that the counts registered the minimum values in each detector. Care was taken to measure the dark counts with the fibers connected to the setup so as to include any thermal radiation persisting in the fiber.

Data was captured at each power level over a time of about 0.5 s, with 2-ns timing resolution. The time tagger gate time was set to 2 ms in the S-fifteen software, which was chosen in order to prevent overflow of the 16 kB memory register and the 32-bit timestamp index at power levels sufficiently large to induce detector saturation.


\section{Analysis Procedure} \label{Analysis Procedure}

Output from the timing electronics, which consisted of a list of detection timestamps and the associated detector triggering pattern, was processed using a script capable of designating a specific coincidence window.  We used a fixed time segment scheme to organize separate coincident window periods and aggregate multiphoton detections in a similar way to the method used in typical PNR detectors to aggregate counts using a gate or pulsed source.  (Another possible method of aggregating multiphoton detections, which we did not use, involves defining coincidence windows relative to individual detections, which is the method that intensity interferometers use to aggregate signals in order to measure $g^{(2)}$ correlations.)  In our case, a fixed software-defined coincidence window schedule is more appropriate for comparison against other PNR detectors, is easier to implement, and avoids issues of repeatedly counting particular detections when grouped among different relatively defined coincidence windows (i.e., overlapping coincidence windows). Moreover, a fixed coincidence schedule is among the coincidence counting schemes that evade the coincidence window loophole for Bell tests \cite{Larsson2014}.

An additional complication, common to each method of aggregating multiphoton counts, is that the detector may reset and register a second detection before the governing coincidence window has expired.  This situation arises when the chosen coincidence window is a large fraction of the detector dead time for free-running SPDs. This problem is not typically seen in most PNR detectors due to the fact that their effective coincidence window tends to be much smaller than the detector dead time.  Source pulsing or detector gating circumvents this complication \cite{Agafonov2007}; however, this was not possible using our setup, as the pulse rise time for our laser source is rather slow (about 5 $\mu$s) and our detectors do not have hardware gating capability.  We avoid this issue in software by forbidding all duplicate detections from the same detector within a coincidence window.  A finite detector dead time also engenders a corresponding complication wherein a possible photon could impinge upon the detector while the detector is in a recovery state.  This complication is common in most PNR detectors. This effect presumably manifests as a drop in detector efficiency and can be mitigated only by improving the detector dead time or decreasing the source intensity.

The script we developed aggregates the multi-detector events, which are typically interpreted as multiphoton events.  An output file specifies the quantities registered for each associated detector combination, which is then fed into another script that processes the multi-detector counts (i.e., click statistics) using a modified binomial detector model based on work by Sperling, Vogel, and Agarwal (SVA) \cite{Sperling2012a,Sperling2012b}.  The binomial model we used to convert click statistics to photon statistics was expanded to permit individualized detector efficiencies and dark counts as well as an unbalanced source distribution across beam splitter paths.  This model was developed for coherent sources and as such is relevant for our laser-based system.  

According to the model, the probability for $k$ clicks is
\begin{equation}
P_{k} = \sum_{|\boldsymbol{m}|=k} \prod_{i=1}^{N} p_i^{m_i} \prod_{j=1}^{N} (1-p_j)^{1-m_j} \; .
\label{eqn:Binomial}
\end{equation}
where $m_i \in \{0,1\}$, $|\boldsymbol{m}| = m_1 + \cdots + m_N$, and $p_i$ is the probability of a detection on the $i^{\rm th}$ detector.  For coherent light, and including detector-specific parameters, $p_i$ is taken to be given by
\begin{equation}
p_i = 1-e^{-\eta_{i}|u_{i}\alpha|^{2}-\nu_{i}} \; ,
\end{equation}
where $\eta_i$ is the detector efficiency, $1-e^{-\nu_i}$ is the probability of a dark count, and $u_i\alpha$ is the amplitude of coherent light entering the detector.  For a uniform beam splitter network, $u_i = 1/\sqrt{N}$.

Nominal initial values for detector efficiencies, dark counts, branch weighting, and the click statistics were input to an optimization routine.  This routine minimizes a chi-squared metric tracking the overall deviation between the model’s binomial photon distribution and the experimental detector click statistics.  For each power level and each coincidence window, the optimization routine was applied to estimate, not only the aforementioned efficiencies, dark counts, and branch weighting, but also the mean photon number, $\mu$. In this way, we could determine the dependence of the mean photon number on both the coincidence window and source power level.

In addition to the experimental data produced, we also developed a script that generates detection events according to a photon-free model with a stochastic vacuum field component combined with deterministic amplitude-threshold detectors \cite{LaCour&Williamson2020,VQOL}. This model has been successfully used to account for quantum effects such as entanglement\cite{LaCourYudichakEI2021} and delayed-choice duality,\cite{LaCourYudichakDC2021}. The classical nature of the model was an intentional feature used to test the possibility of reproducing what is typically interpreted as evidence for the particle nature of light via PNR detectors under purely continuous electromagnetic field conditions.  One could construe this test as an exhibition of the suitability of purely classical approaches for interpreting experimental PNR detector results.

Under this model, coherent light is represented as a complex Gaussian random vector with a nonzero mean.  Specifically, a coherent state $\ket{\alpha}_H\otimes\ket{0}_V$ corresponding to a single spatial mode and two orthogonal polarization modes (horizontal and vertical, respectively) may be represented by the random variables
\begin{subequations}
\begin{align}
a_H &= \alpha + \sigma z_H \\ 
a_V &= \sigma z_V \; ,
\end{align}
\end{subequations}
where $\sigma = 1/\sqrt{2}$ is the standard deviation due to the vacuum state and $z_H, z_V$ are independent standard complex Gaussian random variables.  So $\mathsf{E}[|a_H|^2] - \sigma^2 = |\alpha|^2$ corresponds to the average photon number in the horizontal polarization mode, excluding the vacuum contribution.

For a beam splitter network with $N$ output spatial modes, we now have
\begin{subequations}
\begin{align}
a_{iH} &= u_i \alpha + \sigma z_{iH} \\
a_{iV} &= \sigma z_{iV}
\end{align}
\label{eqn:VQOL}
\end{subequations}
where $u_i$ is defined as before and $z_{iH}, z_{iV}$ are independent standard complex Gaussian random variables.

We treat detections as simple amplitude-threshold-crossing events, so a detection on the $i^{\rm th}$ detector is modeled as the event
\begin{equation}
D_i = \{ |a_{iH}| > \gamma \; \text{or} \; |a_{iV}| > \gamma \} \; .
\label{eqn:detector}
\end{equation}
It is now straightforward to compute the probabilities for various multidetection events, since all detection events are mutually independent.  The probability of exactly $k$ detections will be given by Eqn.\ (\ref{eqn:Binomial}), with $p_i$ replaced by the probability of event $D_i$.


\section{Results} \label{Results}

\begin{figure}[ht]
\centering
\includegraphics[width=\columnwidth]{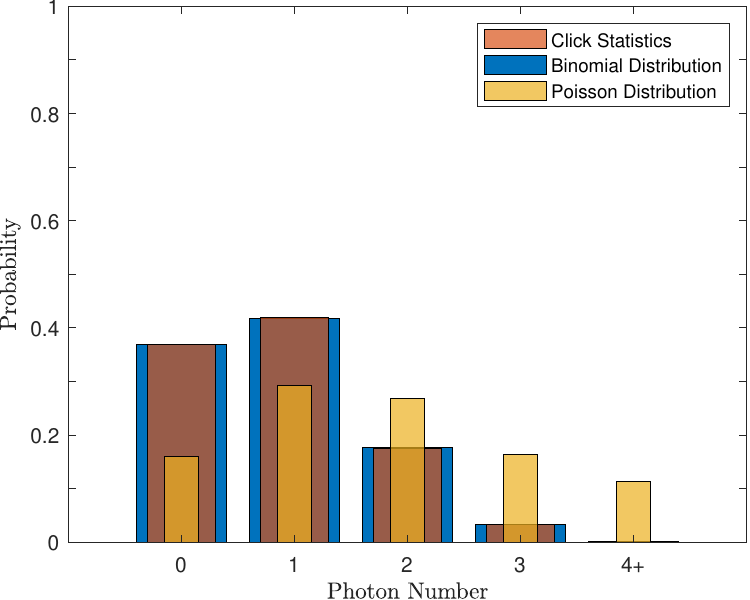}
\caption{(Color online) Distributions describing photon statistics gathered experimentally for a nominal power of 0.11 nW and coincidence windows of 3.5 $\mu$s.  Click statistics, integrated over 0.6 seconds,  are shown in brown.  The fitted binomial distribution is in blue, and the corresponding Poisson distribution, calculated using the binomial model's average photon number, is in gold.  The ``4+'' bin aggregates the probabilities for four or more photons.}
\label{fig:distributions}
\end{figure}

Output from the optimization routine included estimates for the average photon number, detector efficiencies, binomial distribution, and the corresponding Poisson distribution.  The click statistics, binomial distribution, and the corresponding Poisson distribution are plotted in Fig.\ \ref{fig:distributions} for a source power of 0.11 nW and a 3.5 $\mu$s coincidence window resulting in an average photon number of 2.46.  The binary nature of the detectors indicates only the presence or absence of photons \cite{Sperling2013,Bohmann2017,Sperling2014,Piacentini2015,Miatto2018}.  If one were to ascribe to each detection a single captured photon, then many photons would likely be undercounted, as multiple photons may impact a single detector.  Using this na\"{i}ve approach, one would expect a Poisson distribution made manifest in the click statistics directly when the detector system is illuminated by a coherent source.  However, the Poissonian estimation is valid only for extremely low light levels, where the probability of observing a multiphoton state is negligible.  The correct scheme for analyzing arrays of on-off photon detectors is based on the binomial distribution \cite{Sperling2012a,Sperling2012b}.

Instead of using this na\"{i}ve approach, the Poisson distribution in Fig.\ \ref{fig:distributions} was calculated using the average photon number estimated from the optimized binomial distribution.  As such, the Poisson distribution probability weighting is shifted toward higher photon numbers, indicating the veiled undercounting of photons due to multiphoton states producing lower order detector click patterns.  In this way, the Poisson distribution is reconstructed appearing as if single-photon detectors could count multiple photons.  Notably, all probability in the calculated Poisson distribution above a photon number of four is added in the ``4+'' bin, as this PNR system possessed a maximum photon resolution of four corresponding to the number of detectors.  All things considered, Fig.\ \ref{fig:distributions} shows both the agreement between click statistics and the fitted binomial distribution as well as the unsuitability of estimating the photon distribution by fitting a Poisson distribution directly to the click statistics.

\begin{figure}[ht]
\centering
\centerline{(a)}
\includegraphics[width=\columnwidth]{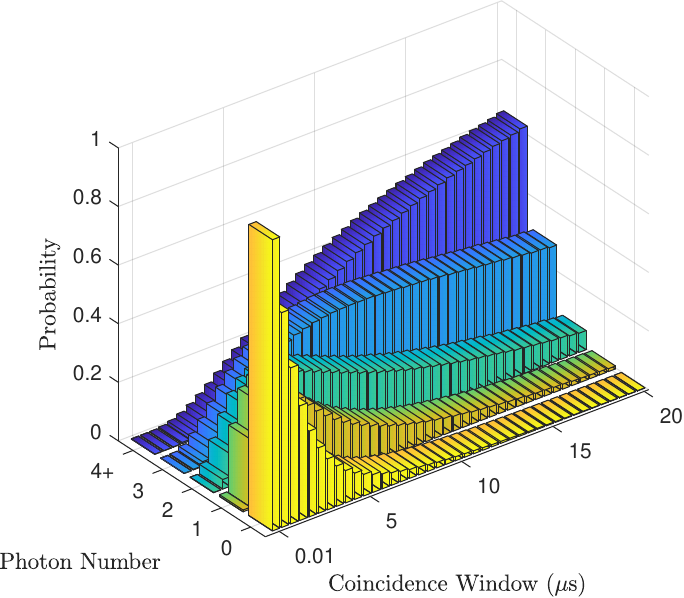}\\
\vspace{0.5cm}
\centerline{(b)}
\includegraphics[width=\columnwidth]{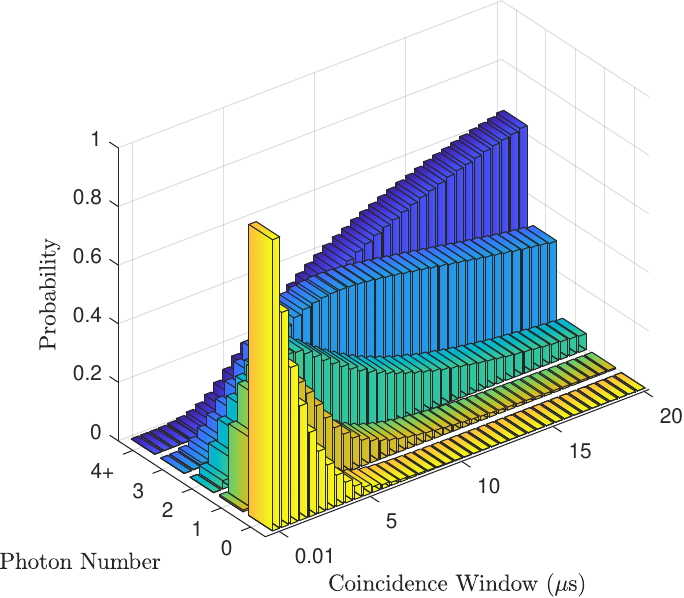}
\caption{(Color online) Top panel (a): Observed click statistics at 0.53 nW as a function of coincidence window.  Bottom panel (b): Model-fitted binomial distribution}
\label{fig:clicks}
\end{figure}

As noted previously, we swept through values of both coincidence window as well as source power.  The top panel of Fig.\ \ref{fig:clicks}, subplot (a), shows the click statistics for all values of coincidence window at a source power level of 0.53 nW, while the bottom panel of Fig.\ \ref{fig:clicks}, subplot (b), shows the corresponding binomial distribution resultant from the modified SVA model. We observe in both cases the weight of the probability shifting from low photon numbers to higher photon numbers as the coincidence window increases. The discrepancy between the observed click statistics and the model-fitted binomial distribution resulted in an average uncertainty below $\pm1.5\%$, showing excellent agreement.
 
\begin{figure}[ht]
\centering
\centerline{(a)}
\includegraphics[width=\columnwidth]{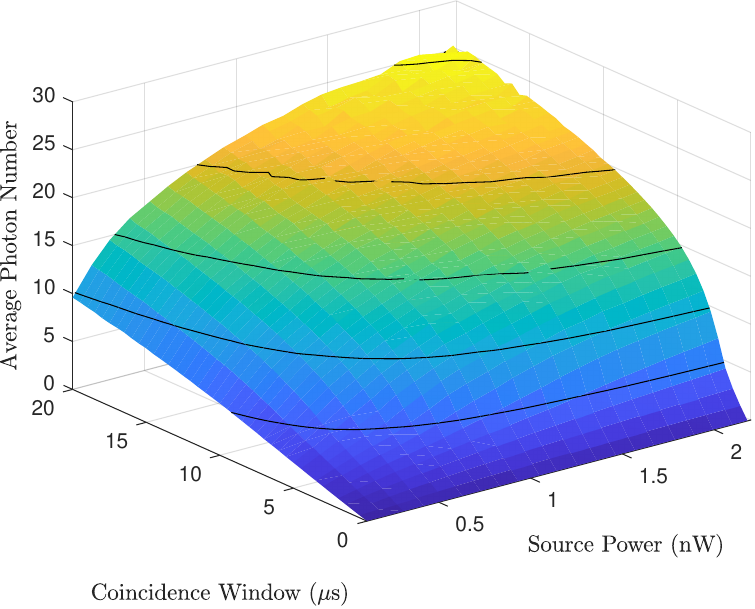}\\
\vspace{0.5cm}
\centerline{(b)}
\includegraphics[width=\columnwidth]{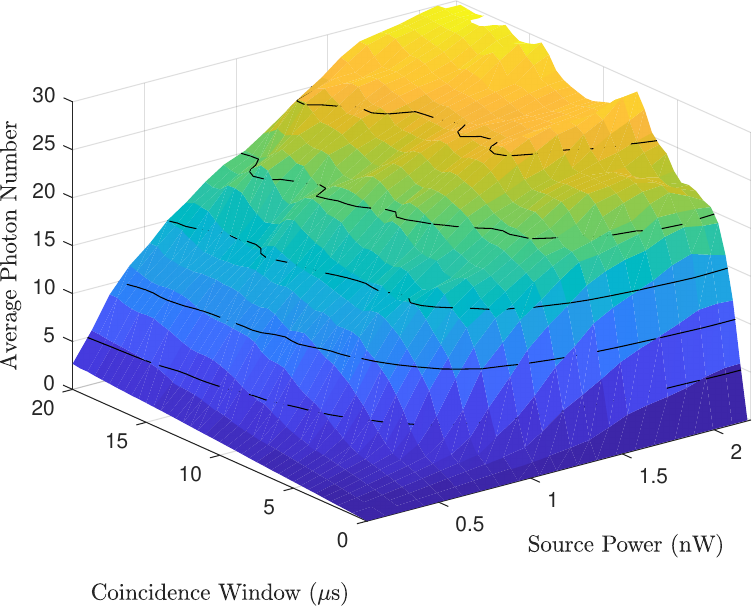}
\caption{(Color online) Top panel (a): Average photon number dependence on coincidence window and source power.  The data was captured experimentally and fitted to the SVA binomial model. Bottom panel (b): Average photon number dependence on coincidence window and source power. This was calculated using a fully classical model of detection events and fitted to the SVA binomial model.}
\label{fig:APN}
\end{figure}

Each binomial photon distribution contained an associated average photon number, which is plotted in the top panel of Fig.\ \ref{fig:APN}, subplot (a), as a function of both the coincidence window and the source power.  We can see that, as expected, the average photon number increases with the source power.  We can also observe the normal saturating effect as the source power induces detection rates on par with the dead time of the detector.  In addition to these expected characteristics, the average photon number behavior also shows a strong dependence on the coincidence window.  To our knowledge no other experimental group have displayed data showing a systematic dependence of calculated photon distribution on coincidence window. The coincidence window dependence, similar to the power dependence, also shows some saturating behavior.  It must be noted, however, that as the average photon number grows above 4.0 the certainty with which we may treat the photon distribution diminishes and consequently the validity of the average photon number begins to rely more heavily on the adherence of the laser’s character as representing a true Poissonian light source.

In addition to gathering experimental data, we also implemented a photon-free model based on stochastic vacuum fluctuations and amplitude threshold detection, as described by Eqns.\ (\ref{eqn:VQOL}) and (\ref{eqn:detector}), in order to test if it is possible to replicate the output of a photon number resolving detector classically, without relying on the \textit{a priori} assumption of light particles. The bottom panel of Fig.\ \ref{fig:APN}, subplot (b), shows the results of this modeling effort in a one-to-one comparison manner with the experimental results. Indeed, the average photon number constructed purely from the results of the classical model and processed in the exact manner as the experimental data shows striking similarity. First, we note the overall similar values on an order-of-magnitude scale, which we emphasize was not guaranteed, as the generally accepted mechanism of operation of PNR detectors, i.e. photon absorption, is not invoked in this simple one-parameter model.  Secondly, we see similar qualitative behavior at comparable values both in the coincidence window timescale as well as the power level. In particular we observe the relatively rapid rise and gradual saturation of average photon number values as each independent variable increases.  As for discrepancies, the model data shows more fine-grained variability even after one application of nearest neighbor averaging, which may be attributed to the slightly shorter overall integration time of the model data. The experimental integration time amounted to 200--300 cycles of 2-ms time periods, providing a total of about half a second of total integration time for each combination of coincidence window and power level.  Likewise, the model data was integrated over 100 periods of 2 ms, which corresponded to similar 10 MB file sizes for experimental data and model data. The file size limit was necessary for prompt processing of the coincidence window script on a PC; i.e., 10 hours total for each model and experiment over all coincidence window and input power combinations. Incidentally, the binomial model optimization script required 11 hours total for each data set.


\section{Discussion} \label{Discussion}

There is an interesting connection between multiplexed PNR detectors and intensity interferometers that may provide some insight into these results.  In fact, intensity interferometers, when operated using Geiger-mode detectors, have exactly the same architecture as a two-detector multiplexed PNR detector.  Furthermore, these similarities persist even if one extrapolates to $N$ detectors.  Not only is the layout topologically equivalent, but the technical analysis of intensity interferometers revealing correlations by grouping coincident detections is essentially the same as counting photon numbers by grouping coincident detections in PNR detectors.  This compelling similarity suggests that the prevailing method of calculating SNR for PNR detectors may be neglecting a more subtle source of error due to accidental coincidences.  Intensity interferometers, by contrast, include such errors explicitly.  Thus, it is prudent to question the certainty with which PNR detector results are presented.  It also suggests that certain anomalous characteristics, such as coincidence window dependent photon statistics, may manifest if accidental coincidences do, in fact, dominate PNR detector results.


\subsection{Theory of Intensity Interferometers} \label{Theory of Intensity Interferometers}

The strong connection between intensity interferometers and PNR detectors, and likewise, connections between measurements of higher order correlations and photon number, are well summarized in a review by Laiho \textit{et al.} \cite{Laiho2022}.  The measurement of correlations between detectors has occupied a pivotal position in the field of quantum optic, due in no small part to its crucial role as a powerful method for extracting experimental evidence concerning the degree of optical coherence of a system \cite{Glauber1963a,Glauber1963b,MandelWolf1995}. The work of Hanbury Brown and Twiss largely founded the practice of such correlation measurements \cite{HBT1954,HBT1956a,HBT1956b}. The intensity interferometers used in their work incorporated pairs of detectors to measure $g^{(2)}$ correlations, thereby quantifying the degree of spatial coherence of stellar sources as a function of detector separation. This technique was used to successfully ascertain the angular dimension of the sources.

Generalizations to higher numbers of simultaneous detections and, as such, access to higher order correlations have been explored theoretically by Malvimat \textit{et al.} \cite{Malvimat2014}. They found that the SNR (measured in dB) of a generalized intensity interferometer using $N$ Geiger-mode single-photon detectors scales as
\begin{equation}
10^{\mathrm{SNR}/10} \sim \frac{N(N-1)}{2} (\eta r\Delta t)^{N/2} \frac{\Delta\tau}{\Delta t} \sqrt{\frac{T}{\Delta t}} \; ,
\label{eqn:SNR}
\end{equation}
where $\eta$ is the detector efficiency, $r$ is the detection rate, $\Delta t$ is the reciprocal electrical bandwidth or, equivalently, the coincidence window, $\Delta \tau$ is the source coherence time, and $T$ is the integration time. Note that Malvimat \textit{et al.}'s original expression did not include detector efficiency, which we have inserted in accordance with Hanbury Brown's SNR expression \cite{HBT1968}.  The symbol $\sim$ reflects the lack of precision of the expression due to individual characteristics of a particular measurement setup, such as transmission losses.

Accidental coincidences, as opposed to correlated  coincidences,  constitute  the  most  elusive source  of  error,  as  other  sources  of  error, such as dark counts and afterpulsing, can be compensated for straightforwardly at the  individual  detector  level.  Accidental coincidences take multiple forms, including dark-count-dark-count combinations between constituent detectors or (effective) pixels, dark-count-light-count combinations, and, most saliently, light-count combinations between different coherence time periods but within single coincidence windows; i.e., when $\Delta t > \Delta \tau$.  The accidental coincidences caused by the mismatch between coherence time and coincidence window establish the primary cause contributing to coincidence window dependence of photon statistics shown earlier.  For pulsed light sources, the coherence time can be no greater than the pulse width (the pulse width commonly becomes the coincidence window if the electrical bandwidth is slower than the pulse width), i.e. $\Delta t \geq \Delta \tau$.  This can limit the coherence time and become disadvantageous for SNR.  For free-running detectors, the quantity $r\Delta t$ is almost always less than unity in experiments, since the detector dead time (recovery or reset) is usually at least as long as the coincidence window. The condition $r\Delta t < 1$ implies that the SNR decreases as the number of detectors increases, and this property has discouraged their use in intensity interferometer experiments.  

\begin{figure}[ht]
\centering
\includegraphics[width=\columnwidth]{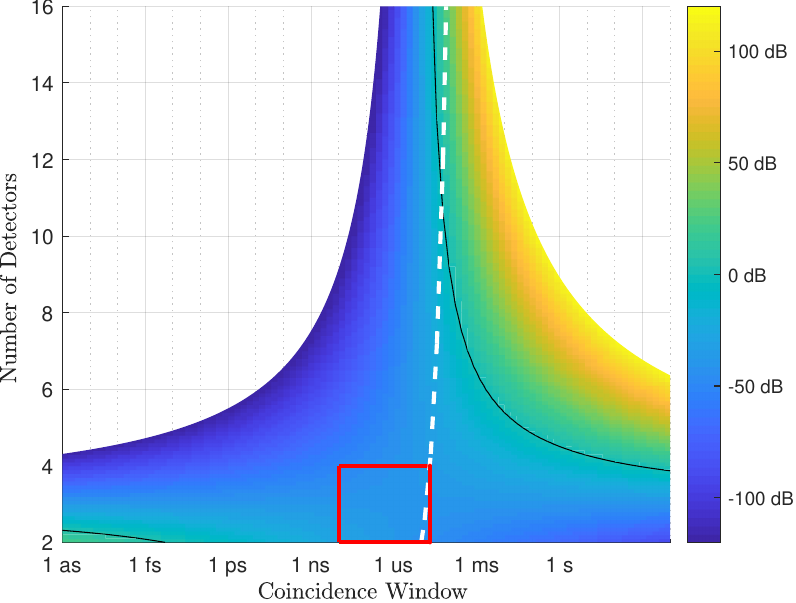}
\caption{(Color online) Theoretical SNR for $N$-detector intensity interferometers. The black contour indicates an SNR of 3 dB. The parameters used in the expression match the experimental conditions used in this work: 2 ps coherence time, 200 kcps detection rate, a 0.5-s integration time. The red box displays the area experimentally explored in this work. The white dashed line represents detector saturation, to the right of which none of the detectors composing the PNR detector are able to fully reset.}
\label{fig:snr}
\end{figure}

From Eqn.\ (\ref{eqn:SNR}), Fig.\ \ref{fig:snr} shows the dependence of SNR on both coincidence window and number of simultaneous detections. The red box represents the parameter space explored experimentally in this work. The white dashed line signifies a limit of effectiveness for PNR detectors beyond which detector saturation spoils the results \cite{Tan2016}. In other words, the number of detections exceeds the reset rate.  Due to saturation, the detector count rate peaks at a maximum count number as input power is increased. The maximum count rate, which characterizes detector temporal saturation, is inversely proportional to the detector dead time. Assuming an equal beam intensity over the surface of a spatially multiplexed PNR detector, all detector segments saturate similarly. In practice, this condition implies that SNR values in the region to the right of the white dashed line are inaccessible due to PNR detector saturation. 

One clear characteristic of Fig. \ref{fig:snr}, owing to the condition $r\Delta t < 1$, is the precipitous falloff of the SNR with increasing detector number. With these experimental parameters, an SNR higher than the 3-dB threshold level (shown by the black line) is reflected only for correlations of order two (which occur at coincidence windows less than 1 fs).  (For reference, Hanbury Brown and Twiss obtained SNRs of 1.2--3 dB in Ref.\ \cite{HBT1956b}.)  Unfortunately, the signal for higher order correlations is overwhelmed by accidental coincidences.  Cognizance of such SNR relations is crucial to avoid mistaking accidental coincidences for correlated coincidences.


\subsection{Review of PNR Detectors} \label{Review of PNR Detectors}

Many existing studies on PNR detectors belie a low SNR when calculated using Eqn.\ (\ref{eqn:SNR}).  Here, we focus on the following most utilized PNR detector architectures: visible light photon counters (VLPCs) \cite{Kim1999, Takeuchi1999, Petroff1989, Baek2010, Waks2004}, transition edge sensors (TESs) \cite{Miller2003, Cabrera1998, Irwin1995, Avella2011, McCammon1984, Gerrits2016, Lita2008, Humphreys2015, Pearlman2010, Levine2012, Namekata2010, Brida2012}, superconducting nanowire single-photon detectors (SNSPDs) \cite{Semenov2001, Goltsman2001, Cahall2017, Zhu2018, Zhang2020, Divochiy2008, Luo2018}, and a class of noncryogenic detectors (NCDs), of which the following sub-types exist: beam splitter single-photon detectors (BS-SPDs) \cite{Achilles2003, Provaznik2020, Hlousek2019}, and spatially \cite{Jiang2007, Avella2016, Ding2019, Kalashnikov2011, Cai2019} or temporally \cite{Kruse2017} multiplexed designs.  In addition, some unique PNR detector designs also exist \cite{Nehra2020, Zambra2005, Straka2018}.
 
\begin{figure}[ht]
\centering
\includegraphics[width=\columnwidth]{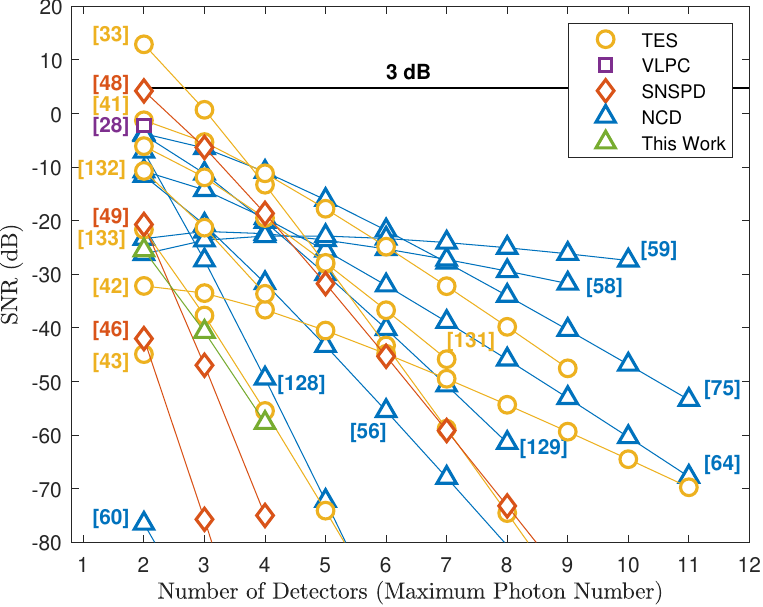}
\caption{(Color online) Survey of theoretical SNR for PNR detectors using Eqn.\ (\ref{eqn:SNR}) and the individually stated device parameters.  Various types of PNR detectors are represented in the survey, including transition edge sensors (TES), visible light photon counters (VLPC), superconducting nanowire single photon detector based devices (SNSPD), and various noncryogenic detectors (NCD).  Lastly, our spatially multiplexed beamsplitter experiment is shown in green.  Few PNR detectors achieve an acceptable SNR, even for a small number of detectors.}
\label{fig:snrsurvey}
\end{figure}

Figure \ref{fig:snrsurvey} shows that SNR typically decreases as a function of increasing detector number.  Three detectors, two TESs and one NCD, manage to achieve an SNR above 3 dB for two-photon events; however, all rapidly decrease, with only one TES remaining above 3 dB for three-photon events.  As one can see, the vast majority of PNR detectors fall below the 3 dB threshold for SNR. At these low SNR values, results would be dominated by accidental coincidences.
 
In addition to the SNR decline, the arbitrary nature of the coincidence window for PNR detectors breeds suspicion with regards to the standard interpretation of photon counting.  In order to fully grasp the importance of arbitrary coincidence windows and their effects on photon statistics, we briefly review some aspects of current PNR detectors.

Simultaneous detection is an ingredient critical to the intended operation of PNR detectors and is vital to probing the veracity of multiphoton states.  In spontaneous parametric down conversion, for example, the decay of one photon in two should be simultaneous; therefore, so should their detection. As noted by Grangier \textit{et al.}, one stipulation of this deduction is that the coincidence window chosen for detection must be no greater than the coherence time of the source \cite{Grangier1986,Grangier2006}. This stipulation ensures that more true coincident detections occur in the detection area rather than accidental, uncorrelated detections. This condition is ostensibly borrowed from SNR concerns when dealing with intensity interferometers. For PNR detectors, if the coherence time is shorter than the coincidence window, we would expect accidental uncorrelated detections to aggregate within coincidence windows and therefore artificially increase the calculated photon number as the coincidence window increases, as was concluded in correlation measurements by Razavi \textit{et al.} \cite{Razavi2009}. This behavior could be tested by varying the coincidence window while keeping the source power constant. In this work, we performed this test by systematically exploring the parameter space comprised of coincidence window duration and input power for a coherent source. However, even if the coincidence window is matched to the coherence time there is still a scaling problem if one increases the number of detectors to capture higher order states, as we have seen in Fig.\ \ref{fig:snr}.

PNR detectors are characterized by several different metrics, such as efficiency, dead time, maximum detectable photon number, photon number resolution, etc. The coincidence window is an often overlooked metric characterizing each particular PNR detector. Coincidence windows are given due attention in loophole-free Bell inequality experiments \cite{Larsson2014,Christensen2015} but are largely neglected for PNR detectors. Obviously, coincidence windows exist in devices regardless of their lack of intentional design in sensor architecture and construction; they are often hardware defined and associated with the slowest response circuitry in the sensor, typically the amplifier system. Oftentimes, the effective coincidence window is merely the reciprocal electrical bandwidth. In this work, we defined the coincidence window as the timescale that determines whether one or more detections should be grouped together, thereby indicating that the detections should be thought of as having a previous association. Already we can see that judgement is an explicit factor in the choice of a coincidence window. In this vein, a software-defined coincidence window, whose only limitation is the hardware circuitry speed, can be tuned to maximize visibility or other metrics of interest \cite{Grangier1986}.

In the case of a hardware-defined coincidence window, the coincidence window is the timescale that characterizes the pileup behavior of the sensor response signal and determines to a large degree how the combination of discrete height pulses in the response signal histogram are distributed. Liao \textit{et al.} \cite{Liao2020} have shown changes in photon statistics while increasing power (and keeping the coincidence window constant). In this work, we varied not only the power of our coherent source but also the coincidence window.

A shift in the photon number distribution caused by merely changing the coincidence window should raise suspicions regarding the accuracy and consistency of the PNR detector in question. Independence of photon number distribution from the coincidence window is regularly and tacitly assumed. As we have shown in this paper, with a beamsplitter tree-based multiplexed quasi-PNR detector 
with admittedly low SNR, the calculated photon number distribution is artificially and strongly dependent upon the coincidence window. In light of this, there is a need for a logically consistent interpretation governing the validity of coincidence window choices with the goal of developing the capability to certify valid PNR results.

In addition to the coincidence window, the coherence time of the light source is pivotal to properly interpreting statistical light distributions measured by PNR detectors.  Lasers and spontaneous parametric down conversion (SPDC) crystals are two of the most common sources for probing the performance characteristics of PNR detectors.  Coherence times for lasers range from 1 ps for entry-level scientific lasers to over 1 ms for high performance, ultra-narrow bandwidth lasers. On the other hand, SPDC source coherence times range from 80 ps for unfiltered output \cite{Halder2008} to 2 $\mu$s for high performance sources \cite{Han2015}.  The atomic cascade source used by Grangier \textit{et al.} for studies on anticorrelation, for example, possessed a lifetime of 4.7 ns \cite{Grangier1986}.  It stands to reason that the stipulation that coincidence windows be no greater than the source coherence time should not only apply to intensity interferometers but also to PNR detectors. 
Experiments have demonstrated that an implicit policy of minimizing the coincidence window seems to be in effect, which would be the correct objective for increasing SNR according to Fig.\ \ref{fig:snr}. However, detector coincidence windows are rarely, if ever, mentioned in comparison to source coherence time, which would be the relevant scaling criterion for determining the suitability of a particular coincidence window. Furthermore, coincidence window minimization across different PNR detector modalities and designs stretches across six orders of magnitude, as shown in Fig.\ \ref{fig:snrsurvey}, increasing the inconsistency with which photon statistics are surmised.

We emphasize that given a coincidence window no greater than the source coherence time, SNR scaling issues persist for higher numbers of detectors and therefore detection of higher order multiphoton states.  To our knowledge no consistent method has been used in these works to establish an optimal coincidence window other than minimization. Since most experimental setups have used the smallest coincidence window available, we are limited only to raising the coincidence window above the hardware defined level, in our case by using a software-defined coincidence window longer than the hardware-defined level of 1 ns.

\begin{figure}[ht]
\centering
\includegraphics[width=\columnwidth]{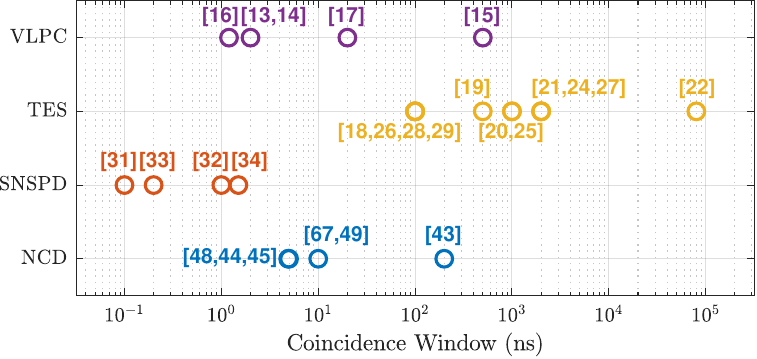}
\caption{(Color online) Survey of coincidence windows for the most relevant PNR detectors including visible light photon counters (VLPC), transition edge sensors (TES), superconducting nanowire single-photon detectors (SNSPD), and noncryogenic detectors (NCD) consisting of spatially and temporally multiplexed single photon detectors and a modified avalanche photodiode.}
\label{fig:CWsurvey}
\end{figure}


\section{Conclusion} \label{Conclusion}

In this work, we have highlighted an error source well known in the field of intensity interferometry, yet, to our knowledge, unrecognized with respect to PNR detectors. This error source comes in the elusive form of accidental coincident detections, which are liable to be mistaken for correlated coincidences.  We discussed the experimental parameters pertinent to establishing sufficient SNR and the importance of the relative values of the coherence time and coincidence window.  Our survey revealed that many common PNR detectors have insufficient SNRs. The PNR detector developed for this work replicated the prevalent low SNR condition to show that variable coincidence windows can indeed alter reported photon statistics.

Under these conditions we were able to establish good agreement between experimental results and a fully classical model based on amplitude threshold detection. It remains to be seen if this classical model is valid for high SNR conditions as well.  If so, it would be possible to attribute the multimodal character of PNR detector output to the nonlinear multiplication amplification processes that are ubiquitous to PNR detectors, which we have modeled as simple threshold exceedances, rather than the detection of discrete physical entities.  We posit this interpretation, which reflects the underlying assumptions of our classical model, as an alternative to the one-to-one causal relationship between a proposed incoming photon and a consequent photoelectron that is more commonly taken as the basis for light quanta. If nothing else this work stands as an appeal to cognizance of the risk of designing instruments that funnel our abilities to observe nature only in terms which we have predesignated.


\ifarxiv
\section*{Acknowledgments}
\else
\begin{acknowledgments}
\fi
This work was supported by the ARL:UT Independent Research and Development Program and by the Office of Naval Research under Grant No.\ N00014-18-1-2107.
\ifarxiv
\else
\end{acknowledgments}
\fi


\ifarxiv
\bibliographystyle{plain}
\else
\bibliographystyle{apsrev4-2}
\fi
\bibliography{refs}

\end{document}